\documentclass[preprint,12pt]{elsarticle}
\usepackage{amsmath}
\usepackage{amssymb}
\usepackage{multirow}
\usepackage{enumerate}
\usepackage{cases}
 \usepackage{amsthm}

\usepackage{mathrsfs}

\makeatletter

\newcommand {\Rmnum} [1] {\expandafter \@slowromancap \romannumeral #1@}
\makeatother

\newtheorem{Remark}{Remark}

\newtheorem{Proposition}{Proposition}

\biboptions{compress}

\begin{document}

\begin{frontmatter}

\title{Mutual authenticated quantum no-key encryption scheme over private quantum channel}


\author{Li Yang$^{1,2}$\corref{1}}
\author{Chenmiao Wu $^{1,2,3}$}
\cortext[1]{Corresponding author email: yangli@iie.ac.cn}
\address{1.State Key Laboratory of Information Security, Institute of Information Engineering, Chinese Academy of Sciences, Beijing 100093, China\\
2.Data Assurance and Communication Security Research Center,Chinese Academy of Sciences, Beijing 100093, China\\
3.University of Chinese Academy of Sciences, Beijing, 100049, China}


\begin{abstract}
We realize shamir's no-key protocol via quantum computation of Boolean permutation  and private quantum channel. The quantum no-key (QNK) protocol presented here is one with mutual authentications, and proved to be unconditionally secure. An important property of this protocol is that0 its authentication key can be reused permanently.
\end{abstract}


\end{frontmatter}



\section{Introduction}
   No-key protocol was first proposed by Shamir \cite{6} which can be used to transmit classical messages secretly in public channel without public key or secret key. Shamir's protocol is based on discrete logarithm problem which cannot resist a man-in-the-middle (MIM) attack. The quantum version of no-key protocol based on single-photon rotations  was developed in \cite{7,8}. The security of quantum no-key (QNK) protocol is based on the laws of quantum mechanics, rather than computational hypothesis. Other similar protocols were proposed \cite{9,10,11}. A protocol proposed in \cite{12} with inherent identification is based on quantum computing of Boolean functions which can prevent MIM attack. Ref. \cite{13} proposed a practical quantum no-key protocol with mutual identification, and present a newly attack named unbalance-of-information-source (UIS) attack. A 9-round QNK protocol with data origin authentication which achieves perfect security was constructed  in \cite{14}. Ref. \cite{1,2} are quantum message oriented protocols which is the development of Shannon's one-time-pad encryption scheme in classical cryptography. Ref. \cite{3} presents some development of those quantum one-time pad schemes. In this paper, we propose a QNK protocol based on the algorithm presented in \cite{1,2}
   
\section{Quantum no-key scheme with interactive identification}
\subsection{Private quantum channel}
Ambainis et al. \cite{2} defined PQC with an ancillary quantum state. Suppose $U_k,k=1,2,\cdots,N$ is a set of operations. Each element $U_k$ is a $2^n\times2^n$ unitary matrix. Let the plaintext state be a n-qubit quantum message $\rho$. In the encryption stage, $U_k$ is applied to the quantum state, where $k$ is a secret key. $p_k$ represents the probability of choosing $k$ as secret key.
\begin{equation}
\rho_c=U_k\rho U_k^{\dagger}.
\end{equation}
To decrypt ciphertext, $U_k^{\dagger}$ is applied to $\rho_c$,
\begin{equation}
\rho=U_k^{\dagger}\rho_c U_k.
\end{equation}
Quantum perfect encryption is defined in \cite{2}: for every input state $\rho$, the output state is an ultimately mixed state, that is
\begin{equation}
\sum_k p_k U_k\rho U_k^{\dagger}=\frac{I}{2^n}.
\end{equation}
\cite{2} constructs one perfect encryption by choosing $p_k=\frac{1}{2^{2n}}$, $U_k=X^{\alpha}Z^{\beta} (\alpha,\beta\in\{0,1\}^n)$. Boykin and Roychowdhury prove that their construction is perfect.

\subsection{Scheme description}
Alice and Bob preshare bit strings $s$ and $r$, $s\in\{0, 1\}^{n}$, $r\in\{0, 1\}^{\frac{n}{2}}$. Alice intends to transmit classical message $x$ to Bob through quantum channel.
\begin{enumerate}
\item Alice randomly selects $\alpha_A,\beta_A \in \{0,1\}^n$ to encrypt $|x\rangle_{\Rmnum{1}}\langle x|$ with $Y^{\alpha_A}H^{\beta_A}$:
\begin{eqnarray}
Y^{\alpha_A}H^{\beta_A}|x\rangle_{\Rmnum{1}}\langle x|H^{\beta_A}Y^{\alpha_A}=\sum\limits_{m}\alpha_{m}|m\rangle_{\Rmnum{1}}\langle m|,
\end{eqnarray}
The first register represents the encrytpion of plaintext.

Then Alice does unitary transform $U_{s}$ on the quantum state:
 \begin{eqnarray}
 &&U_{s}\big(\sum\limits_{m}\alpha_{m}|m\rangle_{\Rmnum{1}}\langle m|\otimes|0\rangle_{\Rmnum{2}}\langle0|\big)U^{\dagger}_{s}\nonumber\\
 &=&\sum\limits_{m}\alpha_{m}|m\rangle_{\Rmnum{1}}\langle m|\otimes|F_{s}(m)\rangle_{\Rmnum{2}}\langle F_{s}(m)|,
 \end{eqnarray}
 and uses $r$ and a randomly selected bit string $r_{A}\in\{0,1\}^{\frac{n}{2}}$ to do exclusive-or operation to get:
 \begin{eqnarray}
 \sum\limits_{m}\alpha_{m}|m\rangle_{\Rmnum{1}}\langle m|\otimes|F_{s}(m)\oplus r\|r_{A}\rangle_{\Rmnum{2}}\langle F_{s}(m)\oplus r\|r_{A}|.
 \end{eqnarray}
The second register consists of the identity information about Alice.

Finally Alice sends Bob registers $\Rmnum{1}$, $\Rmnum{2}$.
\item Bob uses preshared $s$ to do the computation:

 \begin{eqnarray}
&&U^{-1}_{s}(\sum\limits_{m}\alpha_{m}|m\rangle_{\Rmnum{1}}\langle m|\otimes|F_{s}(m)\oplus r\|r_{A}\rangle_{\Rmnum{2}}\langle F_{s}(m)\oplus r\|r_{A}|)(U^{-1}_{s})^{\dagger}\nonumber\\
&=&\sum\limits_{m}\alpha_{m}|m\rangle_{\Rmnum{1}}\langle m|\otimes|F_{s}(m)\oplus F_{s}(m)\oplus r\|r_{A}\rangle_{\Rmnum{2}}\langle F_{s}(m)\oplus F_{s}(m)\oplus r\|r_{A}|\nonumber\\
&=&\sum\limits_{m}\alpha_{m}|m\rangle_{\Rmnum{1}}\langle m|\otimes|r\|r_{A}\rangle_{\Rmnum{2}}\langle r\|r_{A}|,
 \end{eqnarray}
 then Bob measures the second register to get the string $r\|r_{A}$, if the first $\frac{n}{2}$ bits are identical with $r$, he accepts that the message comes from Alice; otherwise, he aborts the scheme.

Through verification, Bob randomly selects $\alpha_B,\beta_B \in \{0,1\}^n$, and uses $Y^{\alpha_B}H^{\beta_B}$ to encrypt:
 \begin{eqnarray}
Y^{\alpha_B}H^{\beta_B}(\sum\limits_{m}\alpha_{m}|m\rangle_{\Rmnum{1}}\langle m|)H^{\beta_B}Y^{\alpha_B}=\sum\limits_{m}\alpha^{'}_{m}|m\rangle_{\Rmnum{1}}\langle m|.
 \end{eqnarray}
The first register contains the transmitted plaintext, and Bob will uses the third register to add his identity information.

Bob does transform $U_{s}$:
 \begin{eqnarray}
&&U_{s}(\sum\limits_{m}\alpha^{'}_{m}|m\rangle_{\Rmnum{1}}\langle m|\otimes|0\rangle_{\Rmnum{3}}\langle0|)U^{\dagger}_{s}\nonumber\\
&=&\sum\limits_{m}\alpha^{'}_{m}|m\rangle_{\Rmnum{1}}\langle m|\otimes|F_{s}(m)\rangle_{\Rmnum{3}}\langle F_{s}(m)|,
 \end{eqnarray}
 and uses $r_{A}$ and a randomly selected $r_{B}$ to do exclusive-or operation, the quantum state becomes:
  \begin{eqnarray}
\sum\limits_{m}\alpha^{'}_{m}|m\rangle_{\Rmnum{1}}\langle m|\otimes|F_{s}(m)\oplus r_{A}\|r_{B}\rangle_{\Rmnum{3}}\langle F_{s}(m)\oplus r_{A}\|r_{B}|.
 \end{eqnarray}

 then sends Alice registers $\Rmnum{1}$, $\Rmnum{3}$.
\item Alice uses $s$ to disentangle the registers:
\begin{eqnarray}
&&U^{-1}_{s}(\sum\limits_{m}\alpha^{'}_{m}|m\rangle_{\Rmnum{1}}\langle m|\otimes|F_{s}(m)\oplus r_{A}\|r_{B}\rangle_{\Rmnum{3}}\langle F_{s}(m)\oplus r_{A}\|r_{B}|)(U^{-1}_{s})^{\dagger} \nonumber\\
 &=&\sum\limits_{m}\alpha^{'}_{m}|m\rangle_{\Rmnum{1}}\langle m|\otimes|r_{A}\|r_{B}\rangle_{\Rmnum{3}}\langle r_{A}\|r_{B}|.
 \end{eqnarray}
Afterwards Alice measures the third register, if first part of the result of measurement is equal to $r_{A}$, she accepts the legality of Bob; otherwise, the scheme is aborted.

Through verification, Alice decrypts with $H^{\beta_A}Y^{\alpha_A}$:
 \begin{eqnarray}
H^{\beta_A}Y^{\alpha_A}(\sum\limits_{m}\alpha^{'}_{m}|m\rangle_{\Rmnum{1}}\langle m|)Y^{\alpha_A}H^{\beta_A}=\sum\limits_{m}\alpha^{''}_{m}|m\rangle_{\Rmnum{1}}\langle m|,
 \end{eqnarray}
and uses $s$ to do transform $U_{s}$ as well as $r$, $r_{B}$ to do exclusive-or operation:
 \begin{eqnarray}
&&\sum\limits_{m}\alpha^{''}_{m}|m\rangle_{\Rmnum{1}}\langle m|\otimes|0\rangle_{\Rmnum{4}}\langle 0|\nonumber\\
&\rightarrow&\sum\limits_{m}\alpha^{''}_{m}|m\rangle_{\Rmnum{1}}\langle m|\otimes|F_{s}(m)\oplus r_{B}\|r_{C}\rangle_{\Rmnum{4}}\langle F_{s}(m)\oplus r_{B}\|r_{C}|,
 \end{eqnarray}
 then sends Bob registers $\Rmnum{1}$, $\Rmnum{4}$.
 \item Bob uses $s$ to do $U^{-1}_{s}$ transform to disentangle the registers:
 \begin{eqnarray}
 &&U^{-1}_{s}(\sum\limits_{m}\alpha^{''}_{m}|m\rangle_{\Rmnum{1}}\langle m|\otimes|F_{s}(m)\oplus r_{B}\|r_{C}\rangle_{\Rmnum{4}}\langle F_{s}(m)\oplus r_{B}\|r_{C}|)(U^{-1}_{s})^{\dagger}\nonumber\\
 &=&\sum\limits_{m}\alpha^{''}_{m}|m\rangle_{\Rmnum{1}}\langle m|\otimes| r_{B}\|r_{C}\rangle_{\Rmnum{4}}\langle r_{B}\|r_{C}|.
 \end{eqnarray}
 By measuring register $\Rmnum{4}$, Bob can verify the legitimacy of Alice. He retains $r_{C}$ to replace $r$. So the preshared bit strings between Alice and Bob for the next session are $s$ and $r_{C}$.

 If Bob makes sure that the message sender is Alice, he decrypts with $H^{\beta_B}Y^{\alpha_B}$:
  \begin{eqnarray}
H^{\beta_B}Y^{\alpha_B}(\sum\limits_{m}\alpha^{''}_{m}|m\rangle_{\Rmnum{1}}\langle m|)Y^{\alpha_B}H^{\beta_B}=|x\rangle_{\Rmnum{1}}\langle x|,
 \end{eqnarray}
 finally Bob gets the transmitted message $x$.
\end{enumerate}
\section{Security analysis}
In the first round communication, if the adversary intercept the transmitted message in the quantum channel, the message state for him is:
  \begin{eqnarray}
  \sigma_{1}&=&\sum\limits_{m,s,r,r_{A}}\alpha_{m}|m\rangle_{\Rmnum{1}}\langle m|\otimes|F_{s}(m)\oplus r\|r_{A}\rangle_{\Rmnum{2}}\langle F_{s}(m)\oplus r\|r_{A}|.
\end{eqnarray}
For every given input $m$, $F_{s}(m)$ iterates through all the possible value. So the quantum state $\sum\limits_{s,r,r_{A}}|F_{s}(m)\oplus r\|r_{A}\rangle_{\Rmnum{2}}\langle F_{s}(m)\oplus r\|r_{A}|$ is an ultimately mixed state which has nothing to do with the value of $m$. Part of the ciphertext state: $\sum\limits_{m}\alpha_{m}|m\rangle_{\Rmnum{1}}\langle m|$ is obtained by performing H and Y on the plaintext state. Now, we firstly prove that the following proposition.
\begin{Proposition}
$\{p_k=\frac{1}{2^{2n}},U_k=U_1^{\alpha}U_2^{\beta}, k=(\alpha,\beta),\alpha,\beta\in\{0,1\}^n\}$ is a quantum perfect encryption.
\end{Proposition}
{\bf Proof:}
Since $\{U_1^\alpha U_2^\beta,\alpha,\beta\in\{0,1\}^n\}$ is a complete orthonormal basis, any $n$-qubit state $\rho$ can be represented as a linear combination of these $2^{2n}$ unitary matrixes: $$\rho=\sum_{\alpha,\beta} a_{\alpha,\beta} U_1^\alpha U_2^\beta,$$
where $a_{\alpha,\beta}=tr(\rho U_2^\beta U_1^\alpha)/{2^n}$.

Thus,

\begin{eqnarray*}
\sum_k p_k U_k\rho U_k^\dagger &=& \frac{1}{2^{2n}}\sum_{\gamma,\delta}U_1^\gamma U_2^\delta \rho U_2^\delta U_1^\gamma \\
&=& \frac{1}{2^{2n}}\sum_{\alpha,\beta}a_{\alpha,\beta}\sum_{\gamma,\delta}U_1^\gamma U_2^\delta U_1^\alpha U_2^\beta U_2^\delta U_1^\gamma.
\end{eqnarray*}
From $U_1U_2=-U_2U_1$,we have $U_2^\delta U_1^\alpha=(-1)^{\alpha\cdot\delta}U_1^\alpha U_2^\delta$. Thus, the above formula can be expressed as:
\begin{eqnarray*}
&& \frac{1}{2^{2n}}\sum_{\alpha,\beta}a_{\alpha,\beta}\sum_{\gamma,\delta}(-1)^{\alpha\cdot\delta}U_1^\alpha U_1^\gamma U_2^\delta (-1)^{\beta\cdot\gamma} U_2^\delta U_1^\gamma U_2^\beta\\
&=& \frac{1}{2^{2n}}\sum_{\alpha,\beta}a_{\alpha,\beta}\sum_{\gamma,\delta}(-1)^{\alpha\cdot\delta}(-1)^{\beta\cdot\gamma}U_1^\alpha U_2^\beta.
\end{eqnarray*}
Because $\frac{1}{2^n}\sum_{\gamma\in\{0,1\}^n}(-1)^{\beta\cdot\gamma}=\delta_{\beta,0}$, the above formula is equal to:
$$
\sum_{\alpha,\beta}a_{\alpha,\beta}\delta_{\alpha,0}\delta_{\beta,0}U_1^\alpha U_2^\beta = a_{00}I=\frac{tr(\rho)}{2^n}I=\frac{I}{2^n}.
$$
So, it is a quantum perfect encryption.$\Box$

Similarly, it's easy to prove that $\{p_k=\frac{1}{2^{2n}},U_k=Y^{\alpha}H^{\beta}, k=(\alpha,\beta),\alpha,\beta\in\{0,1\}^n\}$ also forms a PQC. So $\sum\limits_{m}\alpha_{m}|m\rangle_{\Rmnum{1}}\langle m|$ is an ultimately mixed state.

Thus, the message state $\sigma_{1}$ for the adversary is:
 \begin{eqnarray}
  \sigma_{1}&=&\sum\limits_{m}\alpha_{m}|m\rangle_{\Rmnum{1}}\langle m|\otimes\sum\limits_{s,r,r_{A}}|F_{s}(m)\oplus r\|r_{A}\rangle_{\Rmnum{2}}\langle F_{s}(m)\oplus r\|r_{A}|\nonumber\\
  &=&\frac{I}{2^{n}}\otimes\frac{I}{2^{n}}\nonumber\\
&=&\frac{I}{2^{2n}}.
\end{eqnarray}
Since $\sigma_{1}$ is an ultimately mixed state, the adversary cannot acquire anything by measuring it.

In the second round of communication, the transmitted message state becomes:
\begin{eqnarray}
  \sigma_{2}&=&\sum\limits_{m,s,r_{A},r_{B}}\alpha^{'}_{m}|m\rangle_{\Rmnum{1}}\langle m|\otimes|F_{s}(m)\oplus r_{A}\|r_{B}\rangle_{\Rmnum{3}}\langle F_{s}(m)\oplus r_{A}\|r_{B}|
\end{eqnarray}
  Supposed that the adversary is able to intercept it, the quantum state for him is also an ultimately mixed state:
\begin{eqnarray}
  \sigma_{2}&=&\frac{I}{2^{2n}}.
\end{eqnarray}

Similarly, in the third round, the transmitted message state is also an ultimately mixed state:
\begin{eqnarray}
\sigma_{3}&=&\sum\limits_{m}\alpha^{''}_{m}|m\rangle_{\Rmnum{1}}\langle m|\otimes|F_{s}(m)\oplus r_{B}\|r_{C}\rangle_{\Rmnum{4}}\langle F_{s}(m)\oplus r_{B}\|r_{C}|\nonumber\\
&=&\frac{I}{2^{2n}}.
\end{eqnarray}
Above analysis shows that the preshard $s$, $r$ and secret information $x$ will not be disclosed to the attacker. MIM attack is not effective in this protocol. The adversary has no useful method to attack.
\begin{Remark}
\rm{There are many special cases satisfying the conditions of $U_1$ and $U_2$, such as $X$ and $Z$, $X$ and $Y$, $Y$ and $H$, $X$ and $H$. Thus, the following examples are all quantum perfect encryptions.}
\begin{enumerate}
\item{}
\rm{PQC1:$\{p_k=\frac{1}{2^{2n}},U_k=X^{\alpha}Z^{\beta}, k=(\alpha,\beta),\alpha,\beta\in\{0,1\}^n\}$.}
\item{}
\rm{PQC2:$\{p_k=\frac{1}{2^{2n}},U_k=X^{\alpha}Y^{\beta}, k=(\alpha,\beta),\alpha,\beta\in\{0,1\}^n\}$.}
\item{}
\rm{PQC3:$\{p_k=\frac{1}{2^{2n}},U_k=X^{\alpha}H^{\beta}, k=(\alpha,\beta),\alpha,\beta\in\{0,1\}^n\}$.}
\item{}
\rm{PQC4:$\{p_k=\frac{1}{2^{2n}},U_k=Y^{\alpha}H^{\beta}, k=(\alpha,\beta),\alpha,\beta\in\{0,1\}^n\}$.}

\end{enumerate}
\begin{enumerate}

 \item \rm{When we choose the PQC1: $\{p_k=\frac{1}{2^{2n}},U_k=X^{\alpha}Z^{\beta},\alpha,\beta\in\{0,1\}^n\}$ for QNK protocol, it is insecure to transmit classical information. Because X operation is to reverse the bit and the function of Z operation is to shift the phase. Thus the attacker can measure the ciphertext state in the basis $\{|0\rangle,|1\rangle\}$ without breaking it. And because the three ciphertext transmitted between Alice and Bob are $X^{\alpha_A}Z^{\beta_A}|m\rangle$, $X^{\alpha_B}Z^{\beta_B} X^{\alpha_A}Z^{\beta_A}|m\rangle$,$X^{\alpha_B}Z^{\beta_B}|m\rangle$, the attacker can acquire three strings $\alpha_A\oplus m, \alpha_B\oplus\alpha_A\oplus m, \alpha_B\oplus m$ by measuring the three ciphertext. The attacker can computes $\alpha_B$ with the first string and the second string. Then he can computes the message $m$ with the value of $\alpha_B$ and the third string.}

 \item \rm{When choosing the PQC2: $\{p_k=\frac{1}{2^{2n}},U_k=X^{\alpha}Y^{\beta},\alpha,\beta\in\{0,1\}^n\}$ for the quantum no-key protocol, it is also unsafe to transmit classical information for the same reason. In this case, the three ciphers transmitted between Alice and Bob is $X^{\alpha_A}Y^{\beta_A}|m\rangle$,$X^{\alpha_B}Y^{\beta_B} X^{\alpha_A}Y^{\beta_A}|m\rangle$, $X^{\alpha_B}Y^{\beta_B}|m\rangle$, measuring the three ciphers can achieve the three strings $\alpha_A\oplus\beta_A\oplus m, \alpha_B\oplus\beta_B\oplus\alpha_A\oplus\beta_A\oplus m, \alpha_B\oplus\beta_B\oplus m$. The attacker can computes $\alpha_B\oplus\beta_B$ with the first string and the second string. Then he can computes the message $m$ with the value of $\alpha_B\oplus\beta_B$ and the third string.}
\item In PQC3:$\{p_k=\frac{1}{2^{2n}},U_k=X^{\alpha}H^{\beta}, k=(\alpha,\beta),\alpha,\beta\in\{0,1\}^n\}$, $X$ and $Y$ do not satisfy the condition that $X$ and $Y$ should form an orthonormal basis.

\item By using $Y^{\alpha}H^{\beta}$ in the protocol, the message is being encoded into the conjugate coding, and the flaw stated in the above disappears. If using POC1 and POC2, after the classical bits being encoded into computational basis state, it will stay in computational basis state during the exchange in the protocol. It is better to choose the PQC4: $\{p_k=\frac{1}{2^{2n}},U_k=Y^{\alpha}H^{\beta}, k=(\alpha,\beta),\alpha,\beta\in\{0,1\}^n\}$ for the quantum no-key protocol.

\end{enumerate}
\end{Remark}
Next, we take another attack into account. Assume that the adversary intercepts all the transmitted ciphertext during one session between Alice and Bob.
The transmitted ciphertext during the three rounds of communication are:
 \begin{eqnarray*}
  \sigma_{1}&=&\sum\limits_{m,s,r,r_{A}}\alpha_{m}|m\rangle_{\Rmnum{1}}\langle m|\otimes|F_{s}(m)\oplus r\|r_{A}\rangle_{\Rmnum{2}}\langle F_{s}(m)\oplus r\|r_{A}|,\\
\sigma_{2}&=&\sum\limits_{m,s,r_{A},r_{B}}\alpha^{'}_{m}|m\rangle_{\Rmnum{1}}\langle m|\otimes|F_{s}(m)\oplus r_{A}\|r_{B}\rangle_{\Rmnum{3}}\langle F_{s}(m)\oplus r_{A}\|r_{B}|,\\
\sigma_{3}&=&\sum\limits_{m}\alpha^{''}_{m}|m\rangle_{\Rmnum{1}}\langle m|\otimes|F_{s}(m)\oplus r_{B}\|r_{C}\rangle_{\Rmnum{4}}\langle F_{s}(m)\oplus r_{B}\|r_{C}|.
\end{eqnarray*}

The whole quantum state from adversary's viewpoint is:
\begin{eqnarray}
 \sum\limits_{m_{1},m_{2},m_{3}}\sum\limits_{s,r,r_{A},r_{B},,r_{C}}\alpha_{m_{1}}\alpha^{'}_{m_{2}}\alpha^{''}_{m_{3}}|m_{1},m_{2},m_{3}\rangle_{\Rmnum{1}}\langle m_{1},m_{2},m_{3}|\nonumber\\
\otimes|F_{s}(m_{1})\oplus r\|r_{A}F_{s}(m_{2})\oplus r_{A}\|r_{B},F_{s}(m_{3})\oplus r_{B}\|r_{C}\rangle_{\Rmnum{2}}\times\nonumber\\
\times_{\Rmnum{2}}\langle F_{s}(m_{1})\oplus r\|r_{A},F_{s}(m_{2})\oplus r_{A}\|r_{B},F_{s}(m_{3})\oplus r_{B}\|r_{C}|.
\end{eqnarray}
In \cite{14} , the conclusion is that the authentication key cannot be used forever in the QNK protocol with 3 rounds or less than 3 rounds of communication. If we consider the trace distance between the direct product of any two ciphertext among the three transmitted ciphertext in the proposed QNK protocol in Section 2, we cannot have the result that such trace distance is zero for different plaintext and authentication keys $s$, $r$. As a result, we cannot prove the permanent use of authentication keys $s$, $r$. Guaranteed by the no-cloning theorem, the adversary is unable to copy the unknown quantum state transmitted in the channel. The participants involved in the communication process send message with identification. The message without identity information is not send out into the channel. All the three ciphertext cannot be possessed by the adversary at the same time. So, the coefficients $\alpha_{m_{1}}, \alpha^{'}_{m_{2}}, \alpha^{''}_{m_{3}}$ are distributed in different time and space. The product of $\alpha_{m_{1}}, \alpha^{'}_{m_{2}}, \alpha^{''}_{m_{3}}$ is zero. Thus, it's no use in computing the trace distance between the direct product of any two ciphertext among the three transmitted ciphertext. Moreover, it's also no used in demonstrating that the quantum state show in formula 21 is an ultimately mixed state.

\section{Discussion}
QNK protocol cannot resist MIM attack without identification. The QNK protocol based on PQC without identification is as bellow:
\begin{enumerate}
\item{}
Alice encrypts $\rho$ with $Y^{\alpha_A}H^{\beta_A}$, and sends Bob $\rho_1=Y^{\alpha_A}H^{\beta_A}\rho H^{\beta_A}Y^{\alpha_A}$.
\item{}
Bob encrypts $\rho_1$ with $Y^{\alpha_B}H^{\beta_B}$ and sends Alice $\rho_2=Y^{\alpha_B}H^{\beta_B}\rho_1 H^{\beta_B}Y^{\alpha_B}$.
\item{}
Alice decrypts $\rho_2$ with $H^{\beta_A}Y^{\alpha_A}$ and sends Bob $\rho_3=H^{\beta_A}Y^{\alpha_A}\rho_2 Y^{\alpha_A}H^{\beta_A}$.
\item{}
Bob decrypts $\rho_3$ with $H^{\beta_B}Y^{\alpha_B}$ to recover $\rho$.
\end{enumerate}

If attacker Eve intercepts the message $\rho_1$ from Alice, he randomly selects bit strings $\alpha_{E}$ and $\beta_{E}$ to encrypt $\rho_1$ and sends Alice $\rho^{'}_{2}=Y^{\alpha_E}H^{\beta_E}\rho_1 H^{\beta_E}Y^{\alpha_E}$. Alices decrypts $\rho^{'}_{2}$ with $H^{\beta_A}Y^{\alpha_A}$ and sends Eve $\rho^{'}_3=H^{\beta_A}Y^{\alpha_A}\rho^{'}_2 Y^{\alpha_A}H^{\beta_A}$. Eve receives $\rho^{'}_{3}$ and decrypts it with $H^{\beta_E}Y^{\alpha_E}$. Finally, Eve can get message $\rho$ successfully.

In section 2, we add identification into the protocol to resist MIM attack. Preshard information $r$ and $s$ are necessary in identifying the communicators, so the privacy of $r$ and $s$ are important. We use local random string $r_{A}$, $r_{B}$, Boolean permutation $F_{s}(\cdot)$ and quantum entanglement to protect the Alice and Bob's preshared bit strings $r$ and $s$.

Since the plaintext is encrypted by quantum perfect encyrtion transfromation, the ciphertext state is an ultimately mixed which has nothing to do with the plaintext. In the protocol descryption, we take classical message as example. Moreover, the QNK protocol with identificaiton can be used to transmit quantum message.

\section{Conclusions}
Quantum no-key encryption protocols are presented based on quantum perfect encryption. We make use of random bit strings, Boolean permutation and the property of entanglement to ensure protocols' security. This protocol with identification can  resist MIM attack. The security analysis shows that the pieces of ciphertext of the three rounds are all ultimately mixed states, and the authentication keys can be reused permanently.
\section*{Acknowledgement}
This work was supported by the National Natural Science Foundation of China under Grant No.61173157.


\begin{thebibliography}{99}

\bibitem{6} G. J. Menezes, P. C. van Oorschot and S. A. Vanstone, Handbook of Applied Cryptography, Crc Press, Boca Raton, 1997. 
\bibitem{7} L. Yang and L. A. Wu, Transmit Classical and Quantum Information Secretly, arXiv: quant-ph/0203089. 
\bibitem{8} L. Yang, L. A. Wu and S. H. Liu, Proc. SPIE, 4917: 106-111, 2002.
\bibitem{9} Y. Kanamori, S. M. Yoo and A. S. Mohammad, A Quantum No-key Protocol for Secure Data Communication, Proc 43rd ACM SE Conference, ACM Press, New York, 2005.
\bibitem{10} S. Kak, A Three Stage Quantum Cryptography Protocol, Foundations of Physics Letters \textbf{19}(3), 2006.
\bibitem{11} W. H. Kye, C. M. Kim, M. S. Kim and Y. J. Park, Quantum Key Distribution with Blind Polarization Bases, \emph{Phys. Rev. Lett}, \textbf{95}(4): 040501, 2005.
\bibitem{12} L. Yang, Quantum no-key protocol for direct and secure transmission of quantum and classical messages, arXiv preprint quant-ph/0309200.
\bibitem{13} Y. Wu and L. Yang, Practical quantum no-key protocol with identification, IAS 2009: 540-543, IEEE Computer Society, 2009.
\bibitem{14} L. Yang, Quantum no-key protocol for secure communication of classical message, arXiv:1306.3388, 2013.
\bibitem{1} P. Boykin and V. Roychowdhury, Optimal Encryption of Quantum Bits, \emph{Phys. Rev. A}, \textbf{67}, 042317, 2003.
\bibitem{2} A. Ambainis et al, Private quantum channels,41st Annual Symposium on Foundations of Computer Science, Proceedings:547-553, 2000.
\bibitem{3} A. Nayak and P. Sen, Invertible quantum operations and perfect encryption of quantum states, \emph{QUANTUM INF COMPUT}, \textbf{7}(1-2):103-110, 2007.
\end{thebibliography}

\end{document}